\documentstyle[mprocl,psfig]{article}

\def\beq{\begin{equation}}
\def\eeq{\end{equation}}
\def\vev#1{\langle {#1}\rangle}

\def\be{\begin{equation}}
\def\ee{\end{equation}}

\def\bea{\begin{eqnarray}}
\def\eea{\end{eqnarray}}

\begin{document}

\title{BARYOGENESIS MOTIVATED ON STRING CPT VIOLATION}

\author{O. BERTOLAMI}

\address{Departamento de F\'\i sica, Instituto Superior T\'ecnico,\\
Av. Rovisco Pais, Lisbon, Portugal}

\maketitle\abstracts{We discuss a mechanism for generating the 
baryon asymetry of the Universe that involves a putative violation 
of CPT symmetry arising from string interactions.}

\section{Introduction}

In this contribution we describe a baryogenesis mechanism 
based on a possible violation of the CPT symmetry that arises in string field 
theory \cite{bertolami}. 
Our mechanism is based on the observation that certain string theories may 
spontaneously break CPT symmetry \cite{kp} and Lorentz invariance \cite{ks,ob}.
If CPT and baryon number are violated then
a baryon asymmetry can be generated
in thermal equilibrium \cite{dolgov,cohen}. 
We assume that the source of baryon-number violation 
is due to processes 
mediated by heavy leptoquark bosons of mass 
$M_X$ in a generic GUT whose details are not important in our discussion.

The CPT-violating interactions are shown to arise 
from the trilinear vertex of non-trivial solutions of the field theory 
of open strings and in the corresponding low-energy four-dimensional 
effective Lagrangian via couplings between Lorentz tensors $N$ 
and fermions \cite{kp}.
The CPT and Lorentz invariance violation 
appears when components 
of $N$ acquire non-vanishing vacuum expectation values  $\vev{N}$.
For simplicity, we consider here
only the subset of the CPT-violating terms
leading directly to 
momentum- and spin-independent energy shift 
of particles relative to antiparticles
that are diagonal in the fermion fields, $\psi$,
and involve expectation values  
of only the time components of $N$ \cite{bertolami,kp}:
\beq
{\cal L}_{I} = {\lambda  \vev{N} \over M_S^k} 
\overline{\psi} (\gamma^0)^{k+1} (i \partial_0)^k \psi + h.c. + ...
\quad ,
\label{cptbroken}
\eeq
where $\lambda$ is a dimensionles coupling constant and $M_{S}$
a string mass scale which is presumebly close to the Planck scale. 
Since no large CPT violation has been observed,
the expectation value  $\vev{N}$ must be suppressed
in the low-energy effective theory.
The suppression factor is 
some non-negative power $l$ of 
the ratio of the low-energy scale $m_l$ to $M_S$, that is
$\vev{N}=(m_l/M_S)^l M_S$. 
Since each factor of $i\partial_0$
also provide a low-energy suppression,
the condition $k+l = 2$ corresponds to the dominant terms 
\cite{kp}.
Assuming that each fermion represents 
a standard-model quark of mass $m_q$ and baryon number $1/3$, then 
the energy splitting between a quark and its antiquark
arising from Eq. (1)
can be viewed as an effective chemical potential,
\beq
\mu \sim \left({m_l \over M_S}\right)^l {E^k \over M_S^{k-1}}
\quad ,
\eeq
driving the production of baryon number in thermal equilibrium.
 
The equilibrium phase-space distributions 
of quarks $q$ and antiquarks $\bar q$ at temperature $T$ are 
$f_q(\vec p)=(1+e^{(E - \mu)/T})^{-1}$ and
$f_{\bar q}(\vec p)=(1+e^{(E + \mu)/T})^{-1}$,
respectively, where $\vec p$ is the momentum 
and $E = \sqrt{m_q^2 + p^2}$.
If $g$ is the number of internal quark degrees of freedom,
then the difference between the number densities 
of quarks and antiquarks is
\bea
n_q - n_{\bar q} & = &
{g\over(2\pi)^3}
\int d^3 p ~[f_q(\vec p)-f_{\bar q}(\vec p)]
\quad .
\label{barden}
\eea
The contribution to the baryon-number asymmetry 
per comoving volume is given by 
$n_B/s \equiv (n_q - n_{\bar q})/s$, and on its turn
the entropy density $s(T)$ of relativistic particles is given by 
\beq
s(T) ={2\pi^2\over45} g_s(T) T^3
\quad ,
\label{entropy}
\eeq
where $g_s (T)$ is the sum of the number of degrees of freedom
of relativistic bosons and fermions at temperature $T$.

As shown in Ref. [1] it follows from eqs. (3) and (4) 
that each quark generates a contribution to the baryon number 
per comoving volume of
\beq
{n_q - n_{\bar q} \over s} \sim
{45 g \over 2 \pi^4 g_s(T)} I_k(m_q / T)
\quad ,
\label{basym}
\eeq
where 
\beq
I_k(r) = \int_{r}^\infty dx\, 
{x\sqrt{x^2-r^2} \sinh(\lambda_k x^k)\over\cosh x+\cosh(\lambda_k x^k)}
\quad 
\label{ik}
\eeq
and
\beq 
\lambda_k = \left({m_l \over M_S}\right)^l 
\left({T \over M_S}\right)^{k-1}
\quad .
\eeq

The relevant case for baryogenesis is $k=2$ and 
$\lambda_2 = T/M_S$. 
A good estimate of the integral $I_2(m_q/T)$ 
can be obtained by setting $m_q/T$ to zero,
since fermion masses either vanish or are much smaller than 
the decoupling temperature $T_D$ and hence 
$ I_2(m_q/T) \approx I_2(0) \simeq 7 \pi^4 T/15 M_S$.
This yields for six quark flavours
a baryon asymmetry per comoving volume given by \cite{bertolami}
\beq
{n_B \over s} \simeq {3 \over 5} {T \over M_S}
\quad .
\eeq
Therefore for an appropriate value of the decoupling temperature $T_D$,
the observed baryon asymmetry of the Universe $n_B/s \simeq
10^{-10}$, can be obtained provided the interactions 
violating baryon number are still in
thermal equilibrium at this temperature.
In estimating the value of $T_D$, dilution effects
must be taken into account.

A particularly relevant source of baryon asymmetry dilution
are the baryon violating sphaleron transitions.
These processes are unsuppressed
at temperatures above the electroweak phase transition
\cite{kuzmin}.
Assuming the GUT conserves the quantity $B-L$, $B$ and $L$
denoting the total baryon- and lepton-number densities,
sphaleron-induced baryon-asymmetry dilution
occurs when $B-L$ vanishes
\cite{kuzmin1} and hence \cite{bertolami}:
\beq
{n_B \over s} \simeq
\left({m_L \over  T_W}\right)^2  {T_D \over M_S}
\quad .
\label{result}
\eeq
Taking the heaviest lepton to be the tau 
and the freeze-out temperature $T_W$
to be the electroweak phase transition scale,
then the baryon asymmetry generated via GUT and CPT violating processes
is diluted by a factor of about $10^{-6}$.
Thus, the observed value of the baryon asymmetry 
can be reproduced if,
in a GUT model where $B-L=0$ initially,
baryogenesis takes place 
at a decoupling temperature $T_D \simeq 10^{-4} M_S$,
followed by sphaleron dilution \cite{bertolami}.
This value of $T_D$ is shown to be close to the GUT scale
and leptoquark mass $M_X$,
as required for consistency.

In the less interesting case of GUT models where initially $B-L \not = 0$,
as already mentioned, spharelon dilution effects are not important, however 
other mechanisms such as for instance dilaton decay \cite{yoshimura,bbs},
can set the baryon asymmetry (8) to the observed value.

We point out that the decoupling temperature $T_D \simeq 10^{-4} M_S$
is sufficient low for our baryogenesis mechanism 
to be compatible with string-inspired primordial supergravity 
inflationary models \cite{ross,bb}.

\section*{Acknowledgments}
It is a pleasure to thank D. Colladay, V.A. Kosteleck\'y
and R. Potting for collaboration in the project that gave origin to
this contribution.

\end{document}